\documentclass[trackchanges, twocolumn]{aastex7}
\usepackage{threeparttable}
\usepackage{xcolor}

\newcommand{\authorfootnote}[2][\dagger]{%
  \renewcommand{\thefootnote}{}
}
\definecolor{dkblue}{RGB}{0,0,255}

\newcommand{\RE}[1]{{\textcolor{black}{#1}}}

\begin{document}

\title{Multiwavelength Observations of the Apparently Non-repeating FRB~20250316A}

\setcounter{footnote}{2}
\author[0000-0001-5931-2381]{Ye Li$^\dagger$}
\email[show]{yeli@pmo.ac.cn}
\altaffiliation{These authors contributed equally.}
\affiliation{Purple Mountain Observatory, Chinese Academy of Sciences, Nanjing 210023, China}
\author[0000-0002-9615-1481]{Hui Sun}
\altaffiliation{These authors contributed equally.}
\email{hsun@nao.cas.cn}
\affiliation{National Astronomical Observatories, Chinese Academy of Sciences, Beijing 100101, China}
\author{Lei Qian}
\email{lqian@nao.cas.cn}
\affiliation{National Astronomical Observatories, Chinese Academy of Sciences, Beijing 100101, China}
\author[0000-0002-4562-7179]{Dong-Yue Li}
\email{dyli@nao.cas.cn}
\affiliation{National Astronomical Observatories, Chinese Academy of Sciences, Beijing 100101, China}
\author{Yan-Long Hua}
\email{ylhua@pmo.ac.cn}
\affiliation{Purple Mountain Observatory, Chinese Academy of Sciences, Nanjing 210023, China}
\author{Li-Ping Xin$^\dagger$}
\email[show]{xlp@nao.cas.cn}
\affiliation{National Astronomical Observatories, Chinese Academy of Sciences, Beijing 100101, China}
\author[0000-0001-5798-4491]{Cheng-Kui Li$^\dagger$}
\email[show]{lick@ihep.ac.cn}
\affiliation{Key Laboratory of Particle Astrophysics, Institute of High Energy Physics, Chinese Academy of Sciences, Beijing 100049, China.}
\author[0000-0002-8614-8721]{Yi-Han Wang}
\email{wang3697@wisc.edu}
\affiliation{Department of Astronomy, University of Wisconsin, Madison, WI 53706, USA}
\affiliation{Nevada Center for Astrophysics and Department of Physics and Astronomy, University of Nevada, Las Vegas, NV 89154, USA}
\affiliation{Department of Physics and Astronomy, University of Nevada, Las Vegas, NV 89154, USA}
\author{Jia-Rui Niu}
\email{niujiarui@bao.ac.cn}
\affiliation{National Astronomical Observatories, Chinese Academy of Sciences, Beijing 100101, China}
\author{Tian-Rui Sun}
\email{trsun@pmo.ac.cn}
\affiliation{Purple Mountain Observatory, Chinese Academy of Sciences, Nanjing 210023, China}
\author{Zhu-Heng Yao}
\email{zhyao@bao.ac.cn}
\affiliation{National Astronomical Observatories, Chinese Academy of Sciences, Beijing 100101, China}
\author{Jin-Jun Geng}
\email{jjgeng@pmo.ac.cn}
\affiliation{Purple Mountain Observatory, Chinese Academy of Sciences, Nanjing 210023, China}
\author{Chi-Chuan Jin}
\email{ccjin@nao.cas.cn}
\affiliation{National Astronomical Observatories, Chinese Academy of Sciences, Beijing 100101, China}
\affiliation{School of Astronomy and Space Science, University of Chinese Academy of Sciences, Beijing 100049, China}
\affiliation{Institute for Frontier in Astronomy and Astrophysics, Beijing Normal University, Beijing 102206, China}
\author[0000-0003-2177-6388]{Nanda Rea}
\email{rea@ice.csic.es}
\affiliation{Institute of Space Sciences (ICE, CSIC), Campus UAB, Carrer de Can Magrans s/n, 08193, Barcelona, Spain}
\affiliation{Institut d'Estudis Espacials de Catalunya (IEEC), 08860, Castelldefels, Spain} 
\author{Yuan Liu}
\email{liuyuan@bao.ac.cn}
\affiliation{National Astronomical Observatories, Chinese Academy of Sciences, Beijing 100101, China}
\author{Zhi-Chen Pan}
\email{panzc@nao.cas.cn}
\affiliation{National Astronomical Observatories, Chinese Academy of Sciences, Beijing 100101, China}
\affiliation{Guizhou Radio Astronomical Observatory, Guizhou University, Guiyang 550025, China}
\affiliation{Key Laboratory of Radio Astronomy and Technology (Chinese Academy of Sciences), A20 Datun Road, Chaoyang District, Beijing 100101, China}
\affiliation{School of Astronomy and Space Science, University of Chinese Academy of Sciences, Chinese Academy of Sciences, Beijing 100049, China}
\author{Tao An}
\email{antao@shao.ac.cn}
\affiliation{Shanghai Astronomical Observatory, Chinese Academy of Sciences, Shanghai 200030, China}
\author{Vadim Burwitz}
\email{burwitz@mpe.mpg.de}
\affiliation{Max-Planck-Institut fur extraterrestrische Physik, Giessenbachstrasse 1, 85748 Garching, Germany}
\author{Zhi-Ming Cai}
\email{caizm@microsate.com}
\affiliation{Innovation Academy for Microsatellites, Chinese Academy of Sciences, Shanghai 201210, China}
\author{Jin-Huang Cao}
\email{caojinhuang22@mails.ucas.ac.cn}
\affiliation{National Astronomical Observatories, Chinese Academy of Sciences, Beijing 100101, China}
\author{Yong Chen}
\email{ychen@ihep.ac.cn}
\affiliation{Key Laboratory of Particle Astrophysics, Institute of High Energy Physics, Chinese Academy of Sciences, Beijing 100049, China.}
\author[0000-0003-4200-9954]{Hua-Qing Cheng}
\email{hqcheng@nao.cas.cn}
\affiliation{National Astronomical Observatories, Chinese Academy of Sciences, Beijing 100101, China}
\author{Wei-Wei Cui}
\email{cuiww@ihep.ac.cn}
\affiliation{Key Laboratory of Particle Astrophysics, Institute of High Energy Physics, Chinese Academy of Sciences, Beijing 100049, China.}
\author{Hua Feng}
\email{hfeng@ihep.ac.cn}
\affiliation{Key Laboratory of Particle Astrophysics, Institute of High Energy Physics, Chinese Academy of Sciences, Beijing 100049, China.}
\author{Peter Friedrich}
\email{pfriedrich@mpe.mpg.de}
\affiliation{Max-Planck-Institut fur extraterrestrische Physik, Giessenbachstrasse 1, 85748 Garching, Germany}
\author{Da-Wei Han}
\email{dwhan@ihep.ac.cn}
\affiliation{Key Laboratory of Particle Astrophysics, Institute of High Energy Physics, Chinese Academy of Sciences, Beijing 100049, China.}
\author[0000-0002-0779-1947]{Jing-Wei Hu}
\email{hujingwei@nao.cas.cn}
\affiliation{National Astronomical Observatories, Chinese Academy of Sciences, Beijing 100101, China}
\author{Lei Hu}
\email{leihu@andrew.cmu.edu}
\affiliation{McWilliams Center for Cosmology, Department of Physics, Carnegie Mellon University, 5000 Forbes Ave, Pittsburgh, 15213, PA, USA}
\affiliation{Purple Mountain Observatory, Chinese Academy of Sciences, Nanjing 210023, China}
\author{Yu-Xiang Huang}
\email{huangyuxiang@ynao.ac.cn}
\affiliation{Yunnan Observatories, Chinese Academy of Sciences, Kunming 650216, China}
\affiliation{School of Astronomy and Space Science, University of Chinese Academy of Sciences, Chinese Academy of Sciences, Beijing 100049, China}
\author{Shu-Mei Jia}
\email{jiasm@ihep.ac.cn}
\affiliation{Key Laboratory of Particle Astrophysics, Institute of High Energy Physics, Chinese Academy of Sciences, Beijing 100049, China.}
\author{Ji-An Jiang}
\email{jian.jiang@ustc.edu.cn}
\affiliation{Department of Astronomy, University of Science and Technology of China, Hefei, 230026, China;}
\affiliation{School of Astronomy and Space Sciences, University of Science and Technology of China, Hefei, 230026, China}
\affiliation{National Astronomical Observatory of Japan, National Institutes of Natural Sciences, Tokyo 181-8588, Japan}
\author{Bin Li}
\email{binli@pmo.ac.cn}
\affiliation{Purple Mountain Observatory, Chinese Academy of Sciences, Nanjing 210023, China}
\author{Feng Li}
\email{phonelee@ustc.edu.cn}
\affiliation{State Key Laboratory of Particle Detection and Electronics, University of Science and Technology of China, Hefei 230026, China}
\author{Ming Liang}
\email{liangming@gmail.com}
\affiliation{National Optical Astronomy Observatory (NSF’s National Optical-Infrared Astronomy Research Laboratory) 950 N Cherry Ave. Tucson Arizona 85726, USA}
\author{Yi-Fang Liang}
\email{yfliang@pmo.ac.cn}
\affiliation{Purple Mountain Observatory, Chinese Academy of Sciences, Nanjing 210023, China}
\author{Hao Liu}
\email{lhnows@ustc.edu.cn}
\affiliation{State Key Laboratory of Particle Detection and Electronics, University of Science and Technology of China, Hefei 230026, China}
\author{He-Yang Liu}
\email{liuheyang@nao.cas.cn}
\affiliation{National Astronomical Observatories, Chinese Academy of Sciences, Beijing 100101, China}
\author{Hua-Qiu Liu}
\email{liuhq@microsate.com}
\affiliation{Innovation Academy for Microsatellites, Chinese Academy of Sciences, Shanghai 201210, China}
\author{Norbert Meidinger}
\email{nom@mpe.mpg.de}
\affiliation{Max-Planck-Institut fur extraterrestrische Physik, Giessenbachstrasse 1, 85748 Garching, Germany}
\author{Hai-Wu Pan}
\email{panhaiwu@nao.cas.cn}
\affiliation{National Astronomical Observatories, Chinese Academy of Sciences, Beijing 100101, China}
\author{Arne Rau}
\email{arau@mpe.mpg.de}
\affiliation{Max-Planck-Institut fur extraterrestrische Physik, Giessenbachstrasse 1, 85748 Garching, Germany}
\author{Xin-Wen Shu}
\email{xwshu@mail.ahnu.edu.cn}
\affiliation{Department of Physics, Anhui Normal University, Wuhu, Anhui 241002, China}
\author{Chun Sun}
\email{sunchun@nao.cas.cn}
\affiliation{National Astronomical Observatories, Chinese Academy of Sciences, Beijing 100101, China}
\author{Lian Tao}
\email{taolian@ihep.ac.cn}
\affiliation{Key Laboratory of Particle Astrophysics, Institute of High Energy Physics, Chinese Academy of Sciences, Beijing 100049, China.}
\author{Jin-Long Tang}
\email{ioetang@163.com}
\affiliation{Institute of Optics and Electronics, Chinese Academy of Sciences, Chengdu 610209, China}
\author{Zhen Wan}
\email{zhen_wan@ustc.edu.cn}
\affiliation{Department of Astronomy, University of Science and Technology of China, Hefei, 230026, China;}
\affiliation{School of Astronomy and Space Sciences, University of Science and Technology of China, Hefei, 230026, China}
\author{Hai-Ren Wang}
\email{hairenwang@pmo.ac.cn}
\affiliation{Purple Mountain Observatory, Chinese Academy of Sciences, Nanjing 210023, China}
\author{Jian Wang}
\email{wangjian@ustc.edu.cn}
\affiliation{State Key Laboratory of Particle Detection and Electronics, University of Science and Technology of China, Hefei 230026, China}
\affiliation{Institute of Deep Space Sciences, Deep Space Exploration Laboratory, Hefei 230026, China}
\author{Jing Wang}
\email{wj@bao.ac.cn}
\affiliation{National Astronomical Observatories, Chinese Academy of Sciences, Beijing 100101, China}
\author{Yun-Fei Xu}
\email{xuyf@bao.ac.cn}
\affiliation{National Astronomical Observatories, Chinese Academy of Sciences, Beijing 100101, China}
\author{Yongquan Xue}
\email{xuey@ustc.edu.cn}
\affiliation{Department of Astronomy, University of Science and Technology of China, Hefei, 230026, China;}
\affiliation{School of Astronomy and Space Sciences, University of Science and Technology of China, Hefei, 230026, China}
\author{Xuan Yang}
\email{yangxuan@pmo.ac.cn}
\affiliation{Purple Mountain Observatory, Chinese Academy of Sciences, Nanjing 210023, China}
\author{Da-Zhi Yao}
\email{yaodazhi@pmo.ac.cn}
\affiliation{Purple Mountain Observatory, Chinese Academy of Sciences, Nanjing 210023, China}
\author[0000-0001-6747-8509]{Yuhan Yao}
\email{yuhanyao@berkeley.edu}
\affiliation{Miller Institute for Basic Research in Science, 206B Stanley Hall, Berkeley, CA 94720, USA}
\affiliation{Department of Astronomy, University of California, Berkeley, CA 94720-3411, USA}
\author{Wen Zhao}
\email{wzhao7@ustc.edu.cn}
\affiliation{Department of Astronomy, University of Science and Technology of China, Hefei, 230026, China;}
\affiliation{School of Astronomy and Space Sciences, University of Science and Technology of China, Hefei, 230026, China}
\author{Xiao-Fan Zhao}
\email{zhaoxf@ihep.ac.cn}
\affiliation{Key Laboratory of Particle Astrophysics, Institute of High Energy Physics, Chinese Academy of Sciences, Beijing 100049, China.}
\author{Hong-Fei Zhang}
\email{nghong@ustc.edu.cn}
\affiliation{State Key Laboratory of Particle Detection and Electronics, University of Science and Technology of China, Hefei 230026, China}
\author{Jia-Heng Zhang}
\email{jiahengzhang@mails.ccnu.edu.cn}
\affiliation{National Astronomical Observatories, Chinese Academy of Sciences, Beijing 100101, China}
\author{Juan Zhang}
\email{zhangjuan@ihep.ac.cn}
\affiliation{Key Laboratory of Particle Astrophysics, Institute of High Energy Physics, Chinese Academy of Sciences, Beijing 100049, China.}
\author{Mo Zhang}
\email{mzhang@nao.cas.cn}
\affiliation{National Astronomical Observatories, Chinese Academy of Sciences, Beijing 100101, China}
\author{Song-Bo Zhang}
\email{sbzhang@pmo.ac.cn}
\affiliation{Purple Mountain Observatory, Chinese Academy of Sciences, Nanjing 210023, China}
\affiliation{CSIRO Space and Astronomy, Australia Telescope National Facility, PO Box 76, Epping, NSW 1710, Australia}
\author{Wen-Da Zhang}
\email{wdzhang@nao.cas.cn}
\affiliation{National Astronomical Observatories, Chinese Academy of Sciences, Beijing 100101, China}
\author{Xiao-Ling Zhang}
\email{zhangxl@pmo.ac.cn}
\affiliation{Purple Mountain Observatory, Chinese Academy of Sciences, Nanjing 210023, China}
\author{Yong-He Zhang}
\email{zhangyh@microsate.com}
\affiliation{Innovation Academy for Microsatellites, Chinese Academy of Sciences, Shanghai 201210, China}
\author{Yong-Kun Zhang}
\email{ykzhang@nao.cas.cn}
\affiliation{National Astronomical Observatories, Chinese Academy of Sciences, Beijing 100101, China}
\author{Xian-Zhong Zheng}
\email{xzzheng@sjtu.edu.cn}
\affiliation{Tsung-Dao Lee Institute and Key Laboratory for Particle Physics, Astrophysics and Cosmology, Ministry of Education, Shanghai Jiao Tong University, Shanghai, 201210, China}
\author{Yu-Hao Zhu}
\email{zhuyh@nao.cas.cn}
\affiliation{National Astronomical Observatories, Chinese Academy of Sciences, Beijing 100101, China}
\author{Ying-Xi Zuo}
\email{yxzuo@pmo.ac.cn}
\affiliation{Purple Mountain Observatory, Chinese Academy of Sciences, Nanjing 210023, China}
\author{Sheng-Li Sun}
\email{Palm_sun@mail.sitp.ac.cn}
\affiliation{Shanghai Institute of Technical Physics of the Chinese Academy of Sciences}
\author{Jian-Yan Wei}
\email{wjy@nao.cas.cn}
\affiliation{National Astronomical Observatories, Chinese Academy of Sciences, Beijing 100101, China}
\affiliation{School of Astronomy and Space Science, University of Chinese Academy of Sciences, Chinese Academy of Sciences, Beijing 100049, China}
\author{Wei-Wei Zhu$^\dagger$}
\email[show]{zhuww@nao.cas.cn}
\affiliation{National Astronomical Observatories, Chinese Academy of Sciences, Beijing 100101, China}
\affiliation{Institute for Frontiers in Astronomy and Astrophysics, Beijing Normal University, Beijing 102206, China}
\author{Peng Jiang}
\email{pjiang@nao.cas.cn}
\affiliation{National Astronomical Observatories, Chinese Academy of Sciences, Beijing 100101, China}
\affiliation{Guizhou Radio Astronomical Observatory, Guizhou University, Guiyang 550025, China}
\affiliation{Key Laboratory of Radio Astronomy and Technology (Chinese Academy of Sciences), A20 Datun Road, Chaoyang District, Beijing 100101, China}
\affiliation{School of Astronomy and Space Science, University of Chinese Academy of Sciences, Chinese Academy of Sciences, Beijing 100049, China}
\author{Weimin Yuan}
\email{wmy@nao.cas.cn}
\affiliation{National Astronomical Observatories, Chinese Academy of Sciences, Beijing 100101, China}
\affiliation{School of Astronomy and Space Science, University of Chinese Academy of Sciences, Chinese Academy of Sciences, Beijing 100049, China}
\author[0000-0002-6299-1263]{Xue-Feng Wu}
\email{xfwu@pmo.ac.cn}
\affiliation{Purple Mountain Observatory, Chinese Academy of Sciences, Nanjing 210023, China}
\author[0000-0002-9725-2524]{Bing Zhang}
\email{bzhang1@hku.hk}
\affiliation{Department of Physics, University of Hong Kong, Pokfulam Road, Hong Kong, China}
\affiliation{Nevada Center for Astrophysics and Department of Physics and Astronomy, University of Nevada, Las Vegas, NV 89154, USA}
\affiliation{Department of Physics and Astronomy, University of Nevada, Las Vegas, NV 89154, USA}






\begin{abstract}
The physical origin of fast radio bursts (FRBs) remains uncertain. 
Although multiwavelength observations have been widely conducted, only Galactic FRB~20200428D is associated with an X-ray burst from the magnetar SGR J1935+2154.
Here, we present multiwavelength follow-up observations of the nearby bright FRB~20250316A, including the Five-hundred-meter Aperture Spherical radio Telescope (FAST), Einstein Probe (EP) X-ray mission, Chandra X-ray Observatory, Wide Field Survey Telescope (WFST) and Space Variable Object Monitor/Visible Telescope (SVOM/VT). 
\RE{The 13.08-hour FAST follow-up campaign without pulse detection requires an energy distribution flatter than those of well-known repeating FRBs, suggesting that this burst is likely a one-off event.}
A prompt EP follow-up and multi-epoch observational campaign totaling $>$ 100 ks led to the detection of an X-ray source within the angular resolution of its Follow-up X-ray Telescope (FXT, $10^{\prime\prime}$).
A subsequent Chandra observation revealed this source to be offset by $7^{\prime\prime}$ from the FRB position, and established a 0.5-10 keV flux upper limit of $7.6\times 10^{-15}$ $\rm erg\,cm^{-2}\,s^{-1}$ at the FRB position,
corresponding to $\sim 10^{39}$ $\rm erg\,s^{-1}$ at the 40 Mpc distance of the host galaxy NGC~4141. 
These results set one of the most stringent limits on X-ray emission from a non-repeating FRB, disfavoring ultra-luminous X-ray sources (ULXs) as counterparts of apparently one-off FRBs and offering critical insights into afterglow models.
Our study suggests that an arcsecond localization of both the FRB and its potential X-ray counterpart is essential for exploring the X-ray counterpart of an FRB.

\end{abstract}

\section{Introduction} 
\setcounter{footnote}{0}

Fast Radio Bursts (FRBs) are millisecond-duration extragalactic radio transients \citep{Lorimer2007Sci}. Among the nearly one thousand known FRB sources, fewer than 10\% exhibit repeating bursts \RE{\citep{Spitler2016, chime1stcatalog, CHIME2023repeating}}. Although their origin remains unclear, the association between FRB~20200428D and a pair of X-ray pulses from the Milky Way magnetar SGR~1935+2154 \citep{Mereghetti2020ApJ, CHIME2020Natur, Li2021NatAs} reveals that at least low-luminosity FRBs can be produced by magnetars.

While multiwavelength counterparts of typical extragalactic FRBs have been searched for, no conclusive association has been established (see \citealt{Zhangb2024} for a review). The campaigns include well-studied repeating sources, such as FRB~20121102A \citep{Scholz2016}, FRB~20180301A \citep{Laha2022}, FRB~20180916B \citep{Pilia2020,Scholz2020}, FRB~20190520B \citep{Sydnor2025}, FRB~20220912A \citep{Cook2024}, and FRB~20240114A \citep{Eppel2025}. 
\RE{The most stringent limits in X-ray luminosity}
come from the nearby (3.6 Mpc) repeating FRB~20200120E \citep{Pearlman2025}, which exclude X-ray counterparts similar to ultraluminous X-ray sources (ULXs) or the Crab Nebula. However, most bursts from FRB~20200120E are fainter than typical FRBs \citep{Bhardwaj2021, Kirsten2022, Zhangb2024}, \RE{and} its multiwavelength signals may \RE{also} differ from those of typical FRBs. For apparently non-repeating FRBs, constraints are usually much looser, because they rely primarily on telescopes with large fields of view (FOVs) and relatively poor sensitivities, such as AGILE \citep{Verrecchia2021}, Konus-Wind \citep{Ridnaia2024}, AstroSat-CZTI \citep{Waratkar2025}, and Fermi \citep{Principe2023}. Although two apparently non-repeating FRBs, FRB~20190608B and FRB~20200430A, \RE{constrain X-ray luminosities to be $<10^{40}~\mathrm{erg~s^{-1}}$}, the observations occurred nearly two years after the bursts, and both sources are distant, with redshifts $>0.1$ \citep{Eftekhari2023}.

FRB 20250316A was detected by the Canadian Hydrogen Intensity Mapping Experiment (CHIME, \RE{\citealt{ATel17081, FRB20250316Achime}}) at 08:33:50.859038(3) \RE{UTC} on March 16, 2025 (topocentric at 400 MHz). 
The burst is very bright, with a peak flux of $1.2 \pm 0.1$ kJy and a fluence of $1.7 \pm 0.2$ kJy ms. \RE{High-cadence radio follow-up radio observations detected no pulse \citep{Ould-Boukattine2025, FRB20250316Achime}.} 
The initial localization has arcminute uncertainties.
Subsequent localization using the CHIME core and its three outriggers refined the position to (RA, Dec) = (12h09m44.319s, +58d50m56.708s), with 1$\sigma$ positional uncertainties of 68 mas (semi-major axis) and 57 mas (semi-minor axis) at a position angle of $-0.26^{\circ}$ east of north \RE{\citep{ATel17114, FRB20250316Achime}}.
This FRB is positionally associated with the nearby spiral galaxy NGC 4141, at a distance of $37-44$ Mpc based on the Tully-Fisher relation\RE{\citep{Sorce2014NGC4141, Tully2016NGC4141}}. While an infrared counterpart candidate from James Webb Space Telescope observations was reported \citep{FRB20250316AJWST}, \RE{no clear optical counterpart was detected in MMT and Gemini observations \citep{FRB20250316Achime}}. Assuming a host galaxy distance of 40 Mpc and a central frequency of 600 MHz for CHIME, the isotropic energy and luminosity of the FRB are $(2.0\pm0.2)\times10^{39}~\mathrm{erg}$ and $(1.4\pm0.1)\times10^{42}~\mathrm{erg~s^{-1}}$, respectively.

Owing to its proximity and brightness, FRB 20250316A provides an excellent opportunity for multiwavelength counterpart searches and repeatability studies. To achieve this, we conducted radio, X-ray, and optical follow-up observations shortly after its discovery was reported. The structure of this paper is as follows: Section 2 describes the observations, and Section 3 presents the data analysis and results. We discuss the physical implications in Section 4, and conclude with a summary in Section 5.

\begin{figure*}[!hbt]
    \centering
    \includegraphics[width=1.0\textwidth]{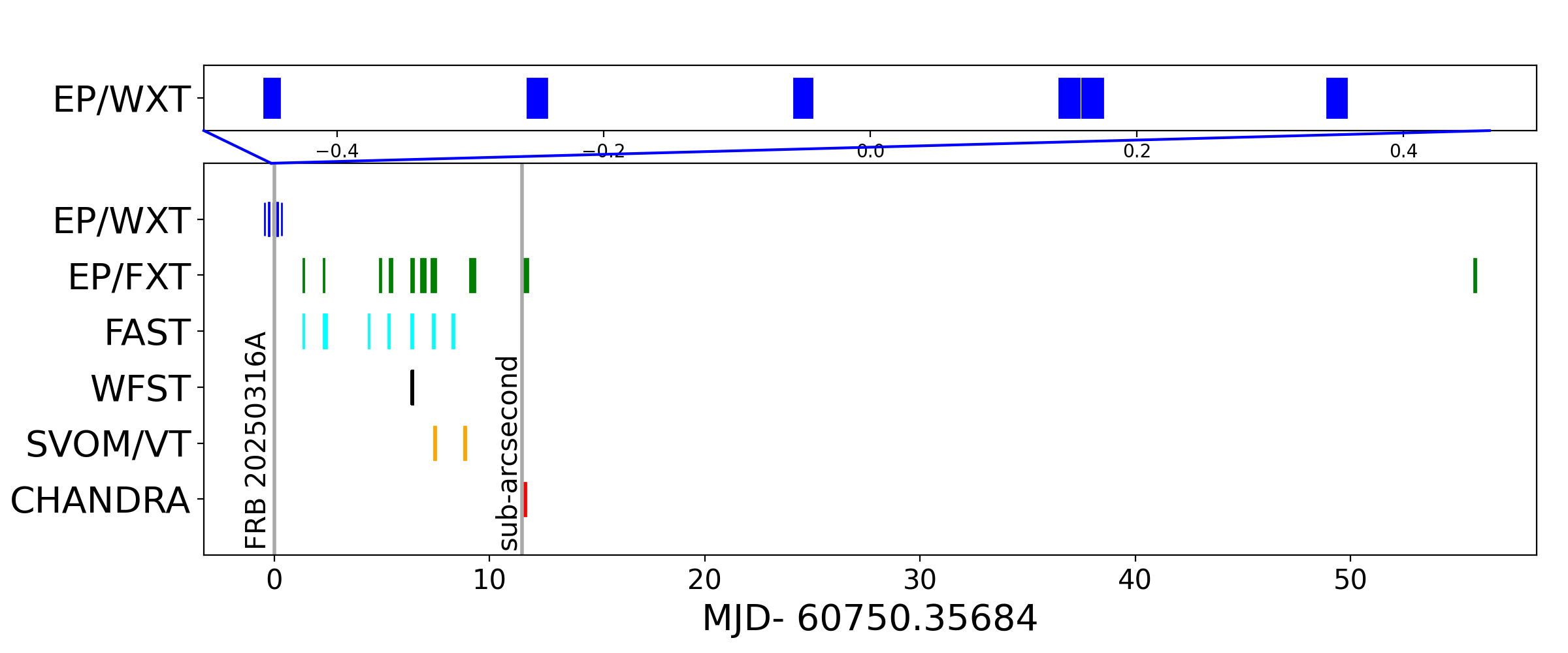}
    \caption{The timeline of the FRB~20250316A follow-up observations. For EP/WXT, only those observations within 12 hours before and after the FRB are presented. Vertical gray lines indicate the time of FRB~20250316A and the time at which FRB~20250316A was localized to be arc-second accuracy. }
    \label{fig:timeline}
\end{figure*}

\section{Observations}

This section summarizes our multiwavelength observations of FRB~20250316A in radio, X-ray, and optical bands. The timeline relative to the FRB is shown in Figure \ref{fig:timeline}.

\subsection{Radio observations}

\begin{table}[!bht] 
\begin{center}
\caption{FAST Observations}
\label{tb:obsfast}
\begin{tabular}{cccc}
\hline \hline
Start Time & MJD & Exposure & Upper limits$^{\rm a}$ \\
  (UTC)   &           & (seconds) & mJy ms \\\hline 
2025-03-17 16:45:00 & 60751.69792 & 3600 & 15.3 \\
2025-03-18 15:38:00 & 60752.65139 & 3600 & 15.3 \\
2025-03-18 16:50:00 & 60752.70139 & 7200 & 15.3 \\
2025-03-20 17:28:00 & 60754.72778 & 3900 & 15.3 \\
2025-03-21 15:39:00 & 60755.65208 & 3900 & 15.3 \\
2025-03-22 16:56:00 & 60756.70556 & 7200 & 15.3 \\
2025-03-23 17:00:00 & 60757.70833 & 7200 & 15.3 \\
2025-03-24 14:58:00 & 60758.62361 & 7200 & 15.3 \\
\hline
\end{tabular}
 $^a$ A burst width of 1 ms was assumed. SNR $>$ 7. 
\end{center}
\end{table}

Following the report of FRB~20250316A \citep{ATel17081}, we triggered the Five-hundred-meter Aperture Spherical radio Telescope (FAST) for follow-up monitoring. Eight sessions between 2025 March 17 and 24 were conducted as part of FAST Director Discretion Time and the FAST FRB key science project \citep{Qian2025} in tracking mode. 
Data from all 19 beams of the FAST L-band receiver ($1.05$–$1.45$ GHz) were channelized into 4096 frequency channels, sampled at 49.153 $\mu$s, and recorded in PSRFITS search mode with dual linear polarization signals 8-bit sampled and processed using ROACH 2. A 1 K noise-switched calibration signal was taken before each observation. The total exposure time was 13.08 hour (13 hours and 5 minutes) across all sessions (Table \ref{tb:obsfast}).

\subsection{X-ray observations}

\subsubsection{Einstein Probe} \label{sec:obs_ep}
The Einstein Probe (EP, \citealt{Yuan2022}) is an X-ray mission launched in January 2024. It carries two instruments: the Wide-field X-ray Telescope (WXT), a monitoring instrument with a field of view of 3600 deg$^2$, and the Follow-up X-ray Telescope (FXT), which provides a large effective area and an angular resolution of $\sim10^{\prime\prime}$. WXT covered the position of FRB~20250316A \RE{in the survey mode} six times within a 24-hour window centered on the burst, with a total exposure of 2 hours (Figure \ref{fig:timeline} and Table \ref{tab:tbxray}). However, the FRB position was not covered at the burst time.
The temporally closest observation started on 2025-03-16 at 07:11:07 UTC and ended at 07:30:49 UTC, about 1 hour before the burst.
This was followed by the first post-burst observation, starting on 2025-03-16 at 11:57:59 UTC, about 3.5 hours after the burst, with an exposure time of 1324 s. 

Following the discovery of FRB 20250316A, we observed the FRB field using the FXT \citep{Chen2021SPIE} onboard EP. Our campaign consisted of ten \RE{target-of-opportunity} observations: nine were taken between 32 hours and 11 days post-burst (Figure \ref{fig:timeline} and Table \ref{tab:tbxray}). An additional late-time observation was carried out on 2025 May 11 to investigate the variability of a previously reported X-ray source in the field, EP J120944.2+585060 \citep{ATel17100,ATel17119}.

\subsubsection{Chandra}
We conducted a follow-up observation of the FRB 20250316A field with the High Resolution Camera (HRC) on the Chandra X-ray Observatory, aiming to search for a potential X-ray counterpart and obtain an arcsecond localization of EP J120944.2+585060. The observation started on 27 March 2025 at 21:31:27 UTC, with an exposure of 10.89 ks, under a Director's Discretionary Time (DDT) proposal (PI: Sun, \dataset [Chandra ObsId 30871]{https://doi.org/10.25574/30871}). The target coordinate was (RA, Dec) (J2000) = (12:09:41.52, +58:50:30.14), the position of EP J120944.2+585060. We selected HRC due to the soft spectrum of EP J120944.2+585060 and its superior effective area below 1 keV compared to other Chandra instruments. The HRC's subarcsecond resolution provides an order-of-magnitude improvement in positional accuracy over the EP-FXT.

\begin{table*}
\begin{center}
\begin{threeparttable}
\caption{X-ray Observations}
\label{tab:tbxray}
\begin{tabular}{cccccc}
\hline \hline
Telescope & Start Time & ObsID & Exposure Time & \multicolumn{2}{c}{Flux($\rm erg\,cm^{-2}\,s^{-1}$)\tnote{a}}\\
 &   (UTC) & & (seconds) & EP J120944.2+585060 & FRB~20250316A\\
\hline
EP-WXT & 2025-03-16 07:11:07 & 11916648071 & 1182 & - &  $< 1.1\times 10^{-11}$ \\
EP-WXT & 2025-03-16 11:57:59 & 11916648077 & 1324 & - &  $< 7.2\times 10^{-12}$ \\
\hline
EP-FXT & 2025-03-17 16:57:29 &  08500000303 & 2994 & - &  $< 3.4\times 10^{-14}$ \\
EP-FXT & 2025-03-18 15:22:38&  08500000304 & 2982 &  $3.4^{+1.0}_{-0.9}\times 10^{-14}$ & - \\
EP-FXT & 2025-03-21 05:50:01&  08500000311 & 5804  & $ 2.9^{+0.7}_{-0.6}\times 10^{-14}$ & - \\
EP-FXT & 2025-03-21 17:02:17&  08500000317 & 10850 & $ 1.6^{+0.4}_{-0.4}\times 10^{-14}$ & - \\
EP-FXT & 2025-03-22 17:03:26&  08500000319 & 10480 & $ 2.3^{+0.5}_{-0.4}\times 10^{-14}$ & - \\
EP-FXT & 2025-03-23 04:15:56&  08500000320 & 16970  &  $ 2.2^{+0.4}_{-0.3}\times 10^{-14}$ & - \\
EP-FXT & 2025-03-23 15:28:24&  08500000321 & 19558 & $ 2.4^{+0.3}_{-0.3}\times 10^{-14}$ & - \\
EP-FXT & 2025-03-25 10:41:53&  08500000325 & 21823 & $ 1.9^{+0.3}_{-0.3}\times 10^{-14}$ & - \\
EP-FXT & 2025-03-27 22:14:34&  08500000326 & 19607 & $ 2.5^{+0.4}_{-0.4}\times 10^{-14}$ & - \\
EP-FXT & 2025-05-11 02:36:50&  06800000592 & 7391 & $ 2.2^{+0.6}_{-0.5}\times 10^{-14}$ & - \\
\hline
Chandra/HRC & 2025-03-27 21:32:27 & 30871 & 10890 & $ 1.6^{+0.5}_{-0.3}\times 10^{-14}$ & $< 7.6\times 10^{-15}$  \\
\hline
\hline
\end{tabular}
\begin{tablenotes}
\item[a] Notes: The unabsorbed X-ray fluxes are given as 90\% upper limits for non-detections at the position of the FRB20250316A. For sources detected within the instrument's positional uncertainty, we report the measured fluxes with their 1$\sigma$ statistical uncertainties. 
\end{tablenotes}
\end{threeparttable}
\end{center}
\end{table*}

\subsection{Optical observations}

\subsubsection{WFST}
The Wide Field Survey Telescope (WFST) is a 2.5-meter aperture facility located in Lenghu, China \citep{wfst}. It features a 6.5 deg$^2$ field of view enabled by its optical design and a mosaic CCD camera. On 2025 March 22, we carried out 21 $r$-band exposures of the FRB~20250316A field, 180 s each, yielding a total integration time of 63 minutes. The start times, exposure durations, and filters of these observations are listed in Table \ref{tb:obswfst} and Table \ref{tb:opt1img}.

\begin{table}[!bht] 
 \setlength{\tabcolsep}{1.5pt} %
\begin{center}
\caption{Optical Observations and Upper Limits}
\label{tb:obswfst}
\begin{tabular}{cccccc}
\hline \hline
Start Time                 & MJD        & T$_\mathrm{exp}$(s)         & Band   & mag$_\mathrm{lim}$       \\\hline\hline
\multicolumn{3}{l}{WFST} & & & \\\hline
\multicolumn{5}{l}{single images} \\
2025-03-22 17:04:19 & 60756.71133 & 180(21) & $r$ & 22.94 - 23.34 \\
\hline
\multicolumn{2}{l}{stacked images} & 180*21 & $r$ & 24.18 \\
\hline\hline
\multicolumn{3}{l}{SVOM/VT} & & & \\\hline
\multicolumn{5}{l}{single images} \\
2025-03-23 18:22:00 & 60757.76458 & 60(101) & VT$\_$B & 21.90 - 22.10 \\
                    &             & 60(101) & VT$\_$R & 21.76 - 22.00 \\
2025-03-25 04:22:05 & 60759.18131 & 60(97) & VT$\_$B & - \\
                    &             & 60(97) & VT$\_$R & - \\
\hline

\multicolumn{2}{l}{stacked images} & 60*101 & VT$\_$B & 24.17 \\
                    &              & 60*101 & VT$\_$R & 24.04 \\

\hline
\hline
\end{tabular}
\end{center}
\end{table}

\subsubsection{SVOM/VT}

The Space-based multi-band astronomical Variable Objects Monitor (SVOM) is a China–France mission designed to detect gamma-ray bursts \citep{svom}. In addition to gamma-ray instruments—GRM (gamma-rays), ECLAIR (hard X-rays), and MXT (soft X-rays)—SVOM carries a 43-cm Visible Telescope (VT) with a dichroic beam splitter dividing light into two channels: VT\_B (0.4–0.65 $\mu$m) and VT\_R (0.65–1 $\mu$m).
SVOM/VT conduected two runs of observations targeting FRB~20250316A, simultaneously acquiring VT\_B and VT\_R data. The first run began 2025-03-23 18:22:00 UTC, the second on 2025-03-25 04:22:05 UTC. Each exposure was 60 s. After removing images affected by background noise or distorted PSFs, 101 and 97 good-quality frames were obtained in the two runs for each band. Details are listed in Tables~\ref{tb:obswfst} and \ref{tb:opt1img}.

\section{Data Analysis and Results}

\subsection{Radio}  \label{sec:res_fast}

Single pulses were searched \RE{for} in the FAST data using PRESTO \RE{\citep{2001PhDT.......123R,2002AJ....124.1788R, 2003ApJ...589..911R}}, with a DM range of 140-180 pc cm$^{-3}$ and a step size of 0.1 pc cm$^{-3}$. No pulses with S/N$>$7 were detected, which corresponds to a fluence threshold of $\sim$15.3 mJy ms for a fiducial burst width of 1 ms and a bandwidth of 400 MHz (Table \ref{tb:obsfast}). This fluence limit translates to an energy upper limit of $1.5 \times 10^{34}$ erg, assuming a source distance of 40 Mpc.  

\subsection{X-ray}
\subsubsection{Einstein Probe}
\begin{figure*}
\centering
\includegraphics[width=0.4\textwidth]{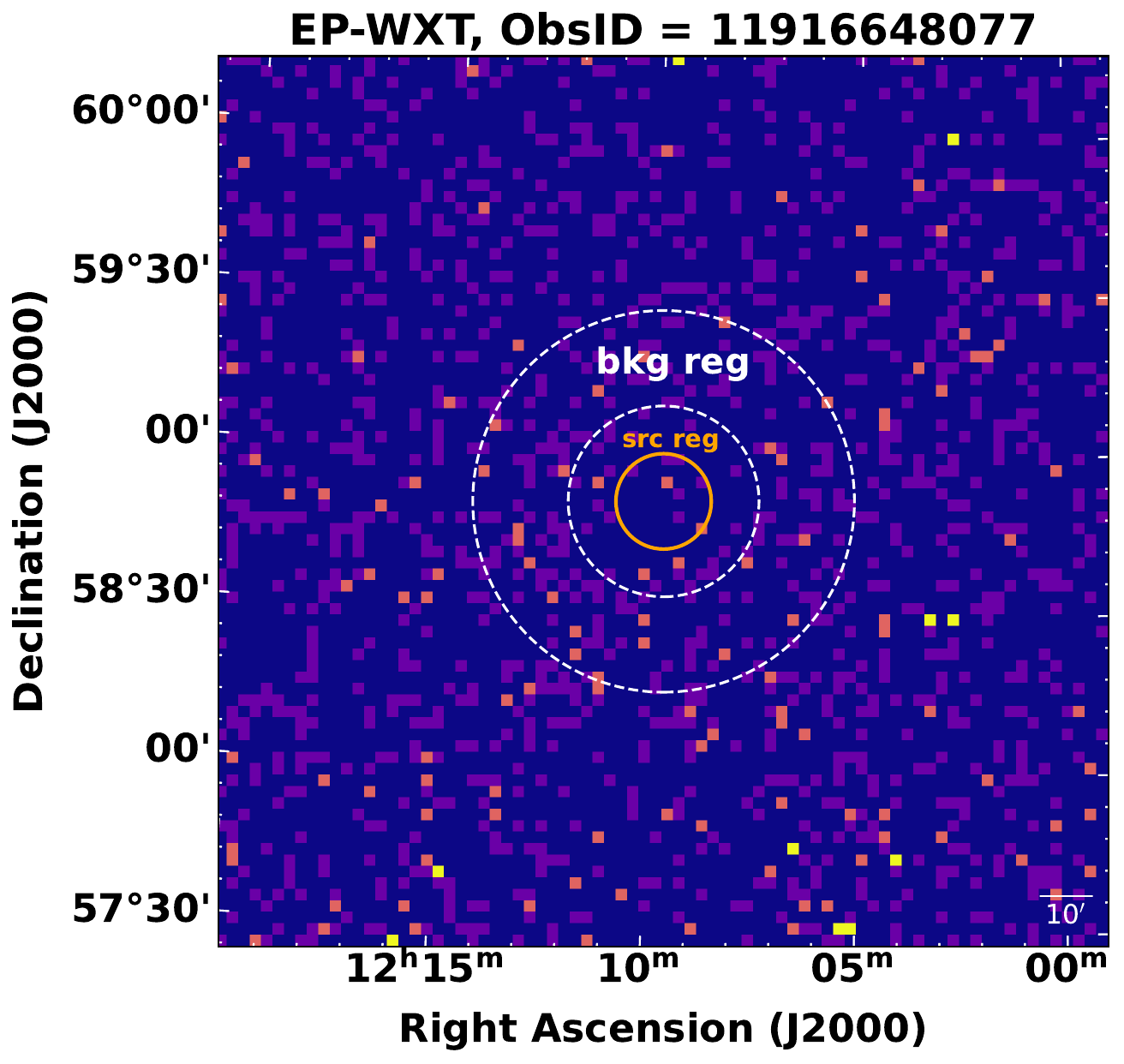}
\includegraphics[width=0.4\textwidth]{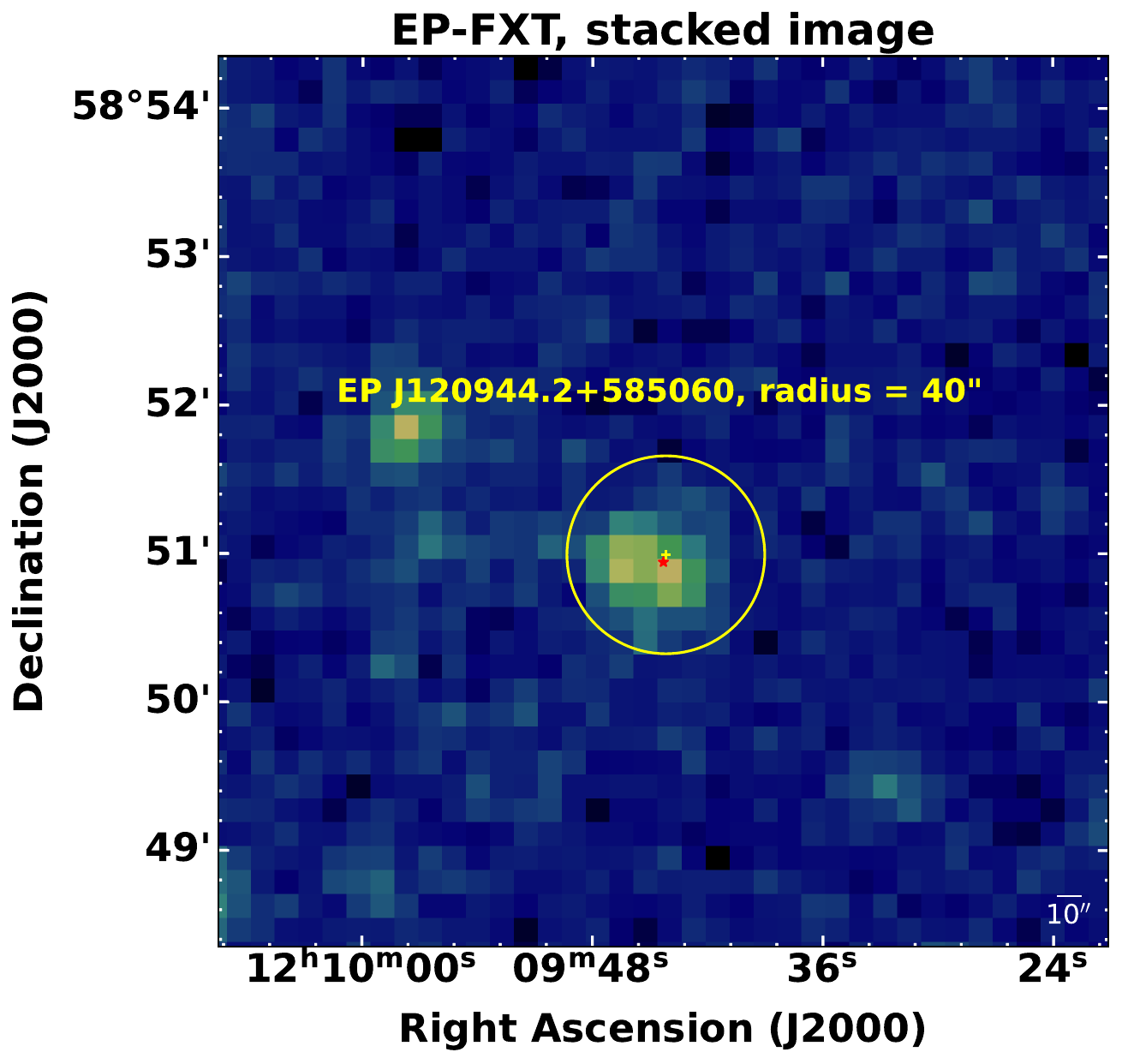}
\includegraphics[width=0.41\textwidth]{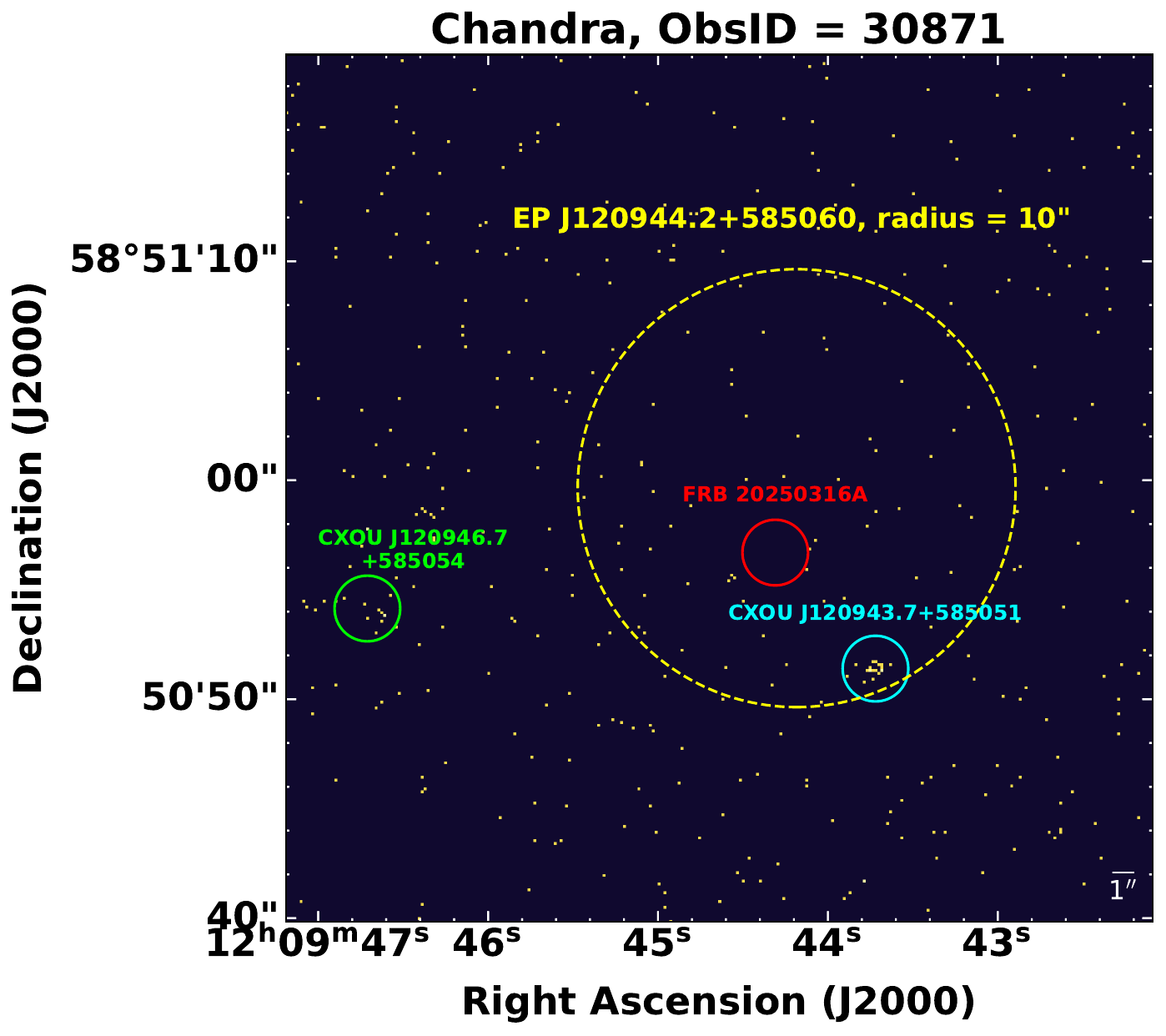}
\includegraphics[width=0.41\textwidth]{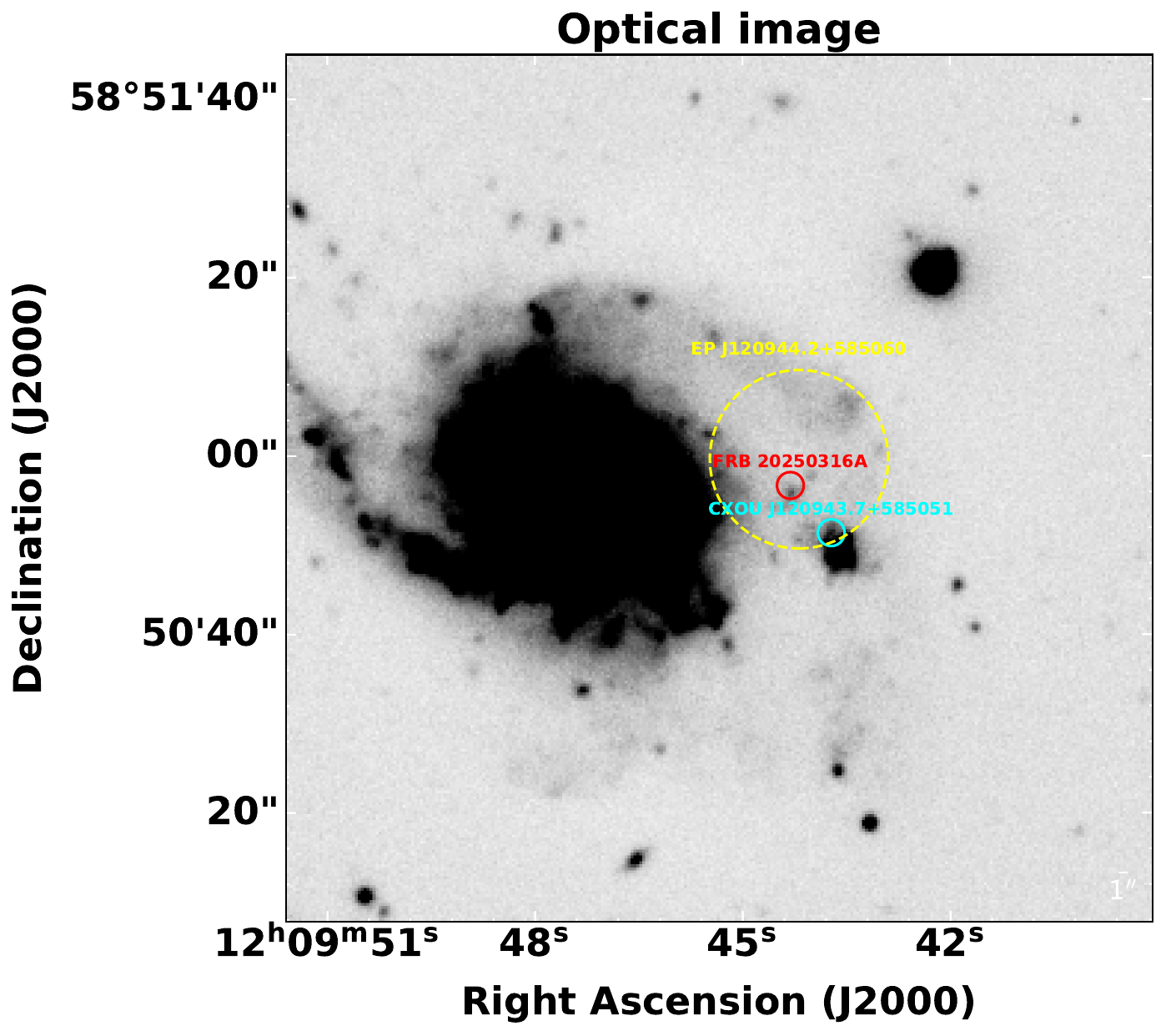}
\caption{X-ray \RE{and optical} images of the FRB 20250316A field.
Upper Left: The EP-WXT image obtained from the observation 3.5 hours after the FRB, with the source and background regions marked for upper limit analysis.  \RE{Upper Right:} The stacked image from all FXT observations, with a yellow circle centered on the position initially reported from the early detection. The red star marks the position of FRB~20250316A.  \RE{Lower Left:} Chandra image. An X-ray source CXOU~J120943.7+585051 (cyan) offset by $\sim 7^{\prime\prime}$ from the FRB position (red circle, projected distance: 1.36 kpc) is located within the error circle of EP J120944.2+585060 (yellow), along with one faint Chandra object (green) at a separation of $\sim 22^{\prime\prime}$ from the first. \RE{Lower Right:} WFST image. Both FRB~20250316A (red) and CXOU~J120943.7+585051 (cyan) reside in separate extended regions. }
\label{fig:x-ray}
\end{figure*}

Using the WXT data reduction pipeline and calibration database \citep[CALDB,][]{Cheng_2025_wxt}, we analyzed the observations obtained 1.5 hours before and 3.5 hours after FRB 20250316A. No significant X-ray sources (above a 5$\sigma$ significance threshold) were detected within the positional uncertainty radius ($\leq$ 3 arcmin) centered at the FRB location (Figure \ref{fig:x-ray}, \RE{upper} left panel). For these two observations, we extracted the source counts from a circular region with a radius of 9 arcmin at the FRB position, \RE{enclosing over 90\% of the photons within the focal spot region as determined by the WXT ground calibration}, and the background counts from an annulus with inner and outer radii of 18 arcmin and 36 arcmin, \RE{respectively,} as shown in Figure \ref{fig:x-ray}. We derived the 90\% confidence count rate upper limits of 0.0057 counts/s and 0.0036 counts/s, respectively, in the 0.5--4 keV range. Assuming an absorbed power-law spectrum with a photon index of 2 and a Galactic absorption column density of $N_H=1.4\times 10^{20}$ $\rm cm^{-2}$, the flux limits are $1.1\times 10^{-11}$ $\rm erg\,cm^{-2}\,s^{-1}$ and $7.2\times 10^{-12}$ $\rm erg\,cm^{-2}\,s^{-1}$, respectively, corresponding to $2.1\times 10^{42}$ $\rm erg\,s^{-1}$ and $1.4\times 10^{42}$ $\rm erg\,s^{-1}$ assuming a luminosity distance of 40 Mpc. 

In the first FXT observation, which began at 2025-03-17T16:57:29 with an exposure time of 2.9 ks, no X-ray sources were detected within the CHIME error region. In the second observation, starting at 2025-03-18T15:23:50 with the same exposure time of 2.9 ks, FXT detected a weak, uncatalogued X-ray source, designated as EP J120944.2+585060. The position of this source is generally consistent with the \RE{first} refined localization provided by CHIME \citep{ATel17086}. 
The source was observed in all eight subsequent FXT observations. For each observation, we extracted the source using a circular region with a 1-arcmin radius. The background was selected from an annulus with inner and outer radii of 1 arcmin and 3 arcmin, \RE{respectively}, where we excluded contamination from bright sources. The FXT data were processed using the FXT data analysis software (fxtsoftware\;v1.20; http://epfxt.ihep.ac.cn/analysis). The process involved particle event identification, pulse invariant conversion, grade calculation and selection (grade 0--12), bad and hot pixel flag and selection of good time intervals using housekeeping files.

The spectrum extracted from the stacked images of all the FXT observations (Figure \ref{fig:x-ray}, \RE{upper right} panel) was fitted in \textit{XSPEC} \RE{(v12.14.1)} with an absorbed power-law model. The best-fit photon index is $2.5\pm 0.2$ (uncertainties at the $1\sigma$ confidence level), assuming the same Galactic absorption as previously introduced. When modeling the spectrum with the absorption as a free parameter, we found negligible intrinsic absorption. The photon index remains consistent with that obtained from the fixed-absorption fit.

The EP-FXT's angular resolution ($\sim$ 23 arcseconds in half power diameter, HPD) \citep{chen2025development} and localization uncertainty ($\lesssim$ 10 arcseconds) are insufficient to conclusively associate the EP source with FRB 20250316A. Therefore, a Chandra/HRC observation was requested to investigate the potential association of the X-ray source with the FRB.

\subsubsection{Chandra}  \label{sec:obs_chandra}

We used the \emph{Chandra} Interactive Analysis of Observations (CIAO, \citealt{Fruscione2006SPIE}; version 4.17) software package and the \emph{Chandra} Calibration Database (CalDB; version 4.12) to reduce and analyze the Chandra observations. By applying the \emph{vtpdetect} in CIAO for source detection, we identified an X-ray source (CXOU~J120943.7+585051) 
within the FXT error circle (Figure \ref{fig:x-ray}, \RE{lower left} panel). The source is located at  
RA (J2000) = 12:09:43.72,
Dec (J2000) = +58:50:51.4, 
with a positional uncertainty of 0.74 arcsec (68\% confidence level, statistical + systematic).
\RE{Employing a circular source region with a radius of 1 arcsec and a background annulus with inner and outer radii of 3 arcsec and 5 arcsec, respectively, we obtained a detection significance of 8$\sigma$ using the Li-Ma formula \citep{lm1983}. The source count rate was derived using the \emph{srcflux} tool, yielding a measurement of $1.56_{-0.34}^{+0.44} \times 10^{-3}$ counts/s (1$\sigma$ uncertainties, 0.1–10.0 keV).}
Assuming an absorbed power-law model with a photon index $\Gamma = 2.5$ and a hydrogen column density $N_\mathrm{H}=1.4\times 10^{20}$ $\rm cm^{-2}$, we estimated the unabsorbed 0.5--10 keV flux from the count rate using PIMMS \citep{Mukai1993} \RE{using the average spectral response files provided within the tool}. The resulting flux is $1.6_{-0.3}^{+0.5}\times 10^{-14}$ $\rm erg\,cm^{-2}\,s^{-1}$ (1$\sigma$ uncertainties), generally consistent with previous EP-FXT measurements, thereby confirming this source as EP J120944.2+585060. 
The position of this source is offset by 7 arcseconds from the precise localization of FRB~20250316A provided by the full CHIME/FRB Outriggers \citep{FRBCollaboration2025}. \RE{The combined positional uncertainty for the two sources is approximately 0.7 arcsecond (derived from the sub-arcsecond uncertainties of both the FRB and Chandra/HRC positions), meaning the observed spatial separation corresponds to nearly 10$\sigma$. This relatively large spatial separation suggests that EP~J120944.2+585060 is not physically associated with FRB~20250316A.}

We note the presence of another object, CXOU~J120946.7+585054, \RE{detected at \RE{4$\sigma$} confidence level within 1 arcmin of the FRB, at the coordinates of RA (J2000) = 12:09:46.71, Dec (J2000) = +58:50:54.1} (with a positional uncertainty of 0.93 arcsec, 68\% confidence level, statistical + systematic, green circle in the \RE{lower left} panel of Figure \ref{fig:x-ray}). 
It is $22^{\prime\prime}$ away from CXOU J120943.7+585051 and $20^{\prime\prime}$ away from the nominal EP~J120944.2+585060. It is difficult for FXT to resolve these two objects, as shown in the stacked FXT image (\RE{upper right} panel of Figure \ref{fig:x-ray}). \RE{The measured count rate for CXOU~J120946.7+585054, obtained with \emph{srcflux}, is $3.98_{-1.65}^{+2.71} \times 10^{-4}$ counts/s (1$\sigma$ uncertainties, 0.1–10.0 keV). Assuming an absorbed power-law model with a photon index $\Gamma = 2$ and the galactic absorption, we estimated the unabsorbed 0.5--10 keV flux from the count rate using PIMMS to be $6.9_{-2.9}^{+4.7}\times 10^{-15}$ $\rm erg\,cm^{-2}\,s^{-1}$.}  
Thus, the flux of CXOU J120943.7+585051 \RE{measured} from the FXT observations are contaminated by $\sim$ 30\% due to CXOU J120946.7+585054.
This contamination may explain why the FXT fluxes are slightly higher than the Chandra flux, although the difference is not statistically significant.

At the position of FRB 20250316A, we derive a 90\% upper limit on the unabsorbed 0.5 -- 10 keV flux of $7.6\times 10^{-15}$ $\rm erg\,cm^{-2}\,s^{-1}$, assuming an absorbed power-law spectrum with a photon index of 2 and $N_\mathrm{H} = 1.4 \times 10^{20}$  $\rm cm^{-2}$. For a distance of 40 Mpc from the host galaxy, this corresponds to a \RE{0.5-10 keV} X-ray luminosity upper limit of $\sim 10^{39}$ $\rm erg\,s^{-1}$. 

\subsubsection{Properties of CXOU~J120943.7+585051}

\begin{figure}
\centering
\begin{tabular}{c}
\includegraphics[width=0.48\textwidth]{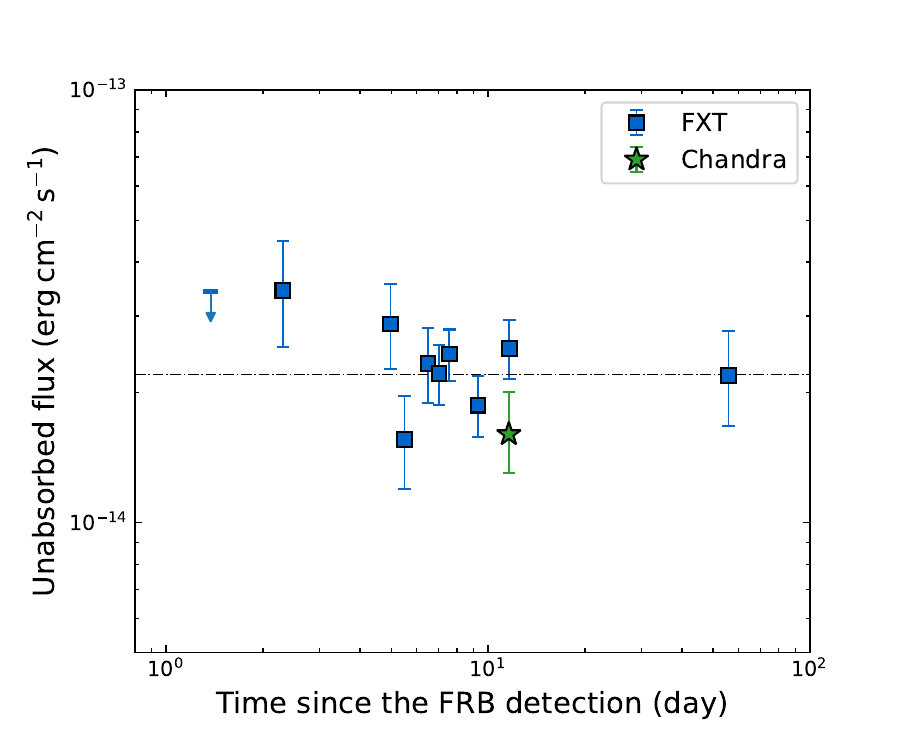}
\end{tabular}
\caption{The X-ray light curve of EP~J120944.2+585060/CXOU~J120943.7+585051 observed by EP-FXT (blue) and \textit{Chandra} (green). Errors show 1$\sigma$ statistical uncertainties. The dashed line shows the weighted mean flux derived from EP-FXT observations. The slightly higher flux measured by EP-FXT may be due to contamination from the nearby \textit{Chandra} source. }
\label{fig:lc_fxt}
\end{figure}

Although the precise location of CXOU~J120943.7+585051/EP~J120944.2+585060 provided by \emph{Chandra} rules out its association with FRB~20250316A, the nature of the X-ray source remains of interest. The source was detected by FXT in observations spanning over 50 days. The $0.5-10$ keV light curve extracted at the Chandra position shows moderate variability 
(see Table \ref{tab:tbxray} and Figure \ref{fig:lc_fxt}). \RE{However, a $\chi^2$ test against a constant flux model yielded $\chi^2 = 8.15$ for 8 degrees of freedom (dof), corresponding to a p-value of 0.419. This high p-value indicates that the observed data are consistent with the null hypothesis of no variability, showing no statistically significant deviation from a constant flux. Together, these results indicate that the flux remained stable within uncertainties across the earlier observations. The late-time observation in May further confirms this steady behavior.}

With an estimated flux of $1.6 \times 10^{-14}$ $\rm erg\,cm^{-2}\,s^{-1}$ inferred from the \emph{Chandra} observation, this source, if associated with the host galaxy at 40 Mpc, corresponds to a luminosity of $\sim 3 \times 10^{39}$ $\rm erg\,s^{-1}$. This places it within the typical luminosity range for ULXs, which are thought to be powered by accretion onto a compact object (either a stellar-mass black hole, neutron star, or potentially an intermediate-mass black hole) at a luminosity comparable to or above the Eddington luminosity of stellar black holes (see \citealt{Kaaret2017ARA&A} for a review). Alternatively, if CXOU J120943.7+585051 is not associated with the FRB's host galaxy, it could be a foreground Galactic source (such as a cataclysmic variable) or a background extragalactic object like an active galactic nucleus (AGN). Distinguishing between these possibilities would require multiwavelength observations.

\subsection{Optical}

\subsubsection{WFST}

WFST images were reduced with standard optical procedures. After extracting objects using SExtractor \RE{\citep{Bertin1996}}, we cross-matched them with Gaia and refined the WCS using SCAMP \RE{\citep{Bertin2006}}. Photometric zero points were calibrated with Pan-STARRS $r$-band magnitudes \RE{\citep{Chambers2016}}. To examine host details, we stacked images using SWarp \RE{\citep{Bertin2002}} with median values.
The stacked image is shown \RE{in the lower right panel of Figure \ref{fig:x-ray}}, with the positions of FRB~20250316A (red), EP~J120944.2+585060 (yellow), and CXOU~J120943.7+585051 (cyan) overplotted. FRB~20250316A and CXOU~J120943.7+585051 lie in two extended regions, likely star-forming regions of the host galaxy.
\RE{The center of the star-forming region hosting FRB~20250316A was determined at (RA, Dec) = (182.434782$^{\circ}$, 58.848778$^{\circ}$) using the {\it centroid\_com} function from Photutils. The FRB lies 1.1$^{\prime\prime}$ from this centroid (215 pc at 40 Mpc), consistent with \cite{FRB20250316Achime} and \cite{Blanchard2025}. To assess systematic uncertainties, we also measured the offset between the FRB and the peak of the star-forming region, yielding 0.8$^{\prime\prime}$ (157 pc).}

After this, we searched for possible \RE{transient} optical counterparts of FRB~20250316A in individual and the stacked image, using the DESI/Legacy Survey (DESI/LS)\footnote{https://www.legacysurvey.org/} $r$ band image as a template. 
Individual images were aligned to the DESI/LS stack with {\it reproject} \citep{Robitaille2020} in Python, then subtracted using HOTPANTS \citep{Becker2015} with a kernel size of $\sim$ 1.5 times the WFST FWHM. No significant objects were found at the FRB~20250316A position in individual subtracted images. The upper limits were estimated as $\rm mag_{lim} = ZP - 2.5\log_{10}( 5\sigma \sqrt{2.226} FWHM)$, where ZP is the zero point of each image and $\sigma$ is the RMS within 90$^{\prime\prime}$ of the FRB in the residual image. The 5$\sigma$ upper limits are listed in Table \ref{tb:obswfst} and Table \ref{tb:opt1img}. An upper limit of $24.18$ mag was estimated from the stacked image using a similar process, limited by the seeing and DESI/LS depth.
\RE{This is generally consistent with the limits of FRB optical counterpart from MMT telescope ($r>25~\mathrm{mag}$ on 2025 March 23) and Gemini ($g>23.7~\mathrm{mag}$ on 2025 March 24) \citep{FRB20250316Achime}.}

\subsubsection{SVOM/VT}

We first stacked the images obtained on 2025-03-23 and 2025-03-25 individually, following the standard procedure using IRAF v2.16~\citep{Tody1986, Tody1993}:
(1) Images with abnormally high backgrounds or distorted stellar profiles were excluded. The remaining images underwent bias, dark, and flat-field corrections.
(2) Precise astrometric alignment was performed with astrometry.net~\citep{Lang2010, joseph_2024_12684908} relative to a chosen reference frame using common bright stars.
(3) The aligned images were then stacked with IRAF.IMCOMBINE task with an average combination method, incorporating sigma-clipping to reject cosmic rays and bad pixels, improving the signal-to-noise ratio of the final stacked images.
For both the VT$\_$B and VT$\_$R bands, we stacked 101 and 97 frames from 2025-03-23 and 2025-03-25, respectively, each with 60 s exposure.
We then searched for possible optical counterparts with a similar process to that of WFST. Because SVOM/VT observations were taken on two different days, and used custom-designed filters ($400-650$ nm for VT$\_$B and $650-1000$ nm for VT$\_$R), 
we utilized the images from SVOM/VT itself as templates to perform image subtraction. Since signals temporally closer to the burst are expected to be more significant, the last single and the stacked images from 2025-03-25 were used as templates, for single and stacked image subtraction, respectively. The upper limits are summarized in Table \ref{tb:obswfst} and Table \ref{tb:opt1img}. 

\section{Physical Implications and Discussions}

\subsection{Probability of Chance Coincidence}

\begin{figure}[!htb]
\centering
\begin{tabular}{c}
\includegraphics[width=0.45\textwidth]{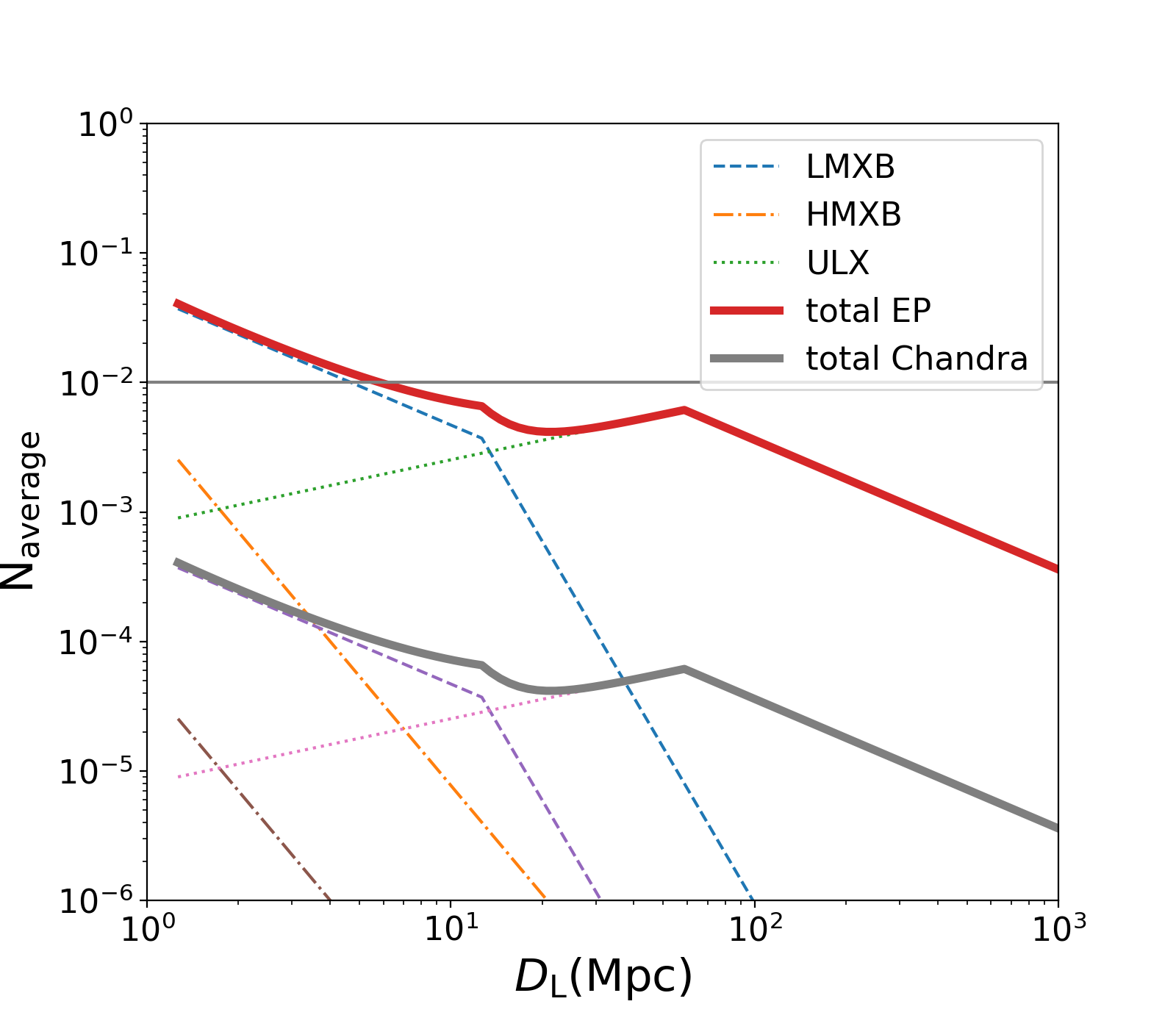}
\end{tabular}
\caption{
\RE{ Averaged number of unrelated X-ray sources within the augular resolution for EP/FXT and Chandra telescope.}}
\label{fig:Nave}
\end{figure}

The X-ray source CXOU~J120943.7+585051 was detected by \textit{Chandra} within the FXT error circle. Given its relatively large separation of 7 arcseconds from the FRB position, the source is unlikely to be physically associated with FRB~20250316A. 
To quantify the likelihood of such a chance alignment, we calculated the probability of 
\RE{chance coincidence in this section.}
It should be noted that alternative origins for CXOU~J120943.7+585051, such as a foreground Galactic X-ray source or a background AGN, cannot be fully ruled out.

\RE{Firstly, we estimate the probability to have an X-ray source within a circle with a 10$^{\prime\prime}$ radius on average. Based on the X-ray source density \citep{Masini2020} at EP/FXT limiting flux, $10^{-14}\,\mathrm{erg\,cm^{-2}\,s^{-1}}$ in $0.5-4$ keV, the averaged number of X-ray sources within the EP/FXT angular resolution is $0.003$. }

\RE{Secondly, the density of X-ray sources within galaxies may be enhanced. Thus, we also calculated the probability of a random, unrelated ULX appearing within 10 arcseconds of the FRB position in a galaxy.}
A 10-arcsecond angular separation at 40 Mpc corresponds to a projected physical radius of 1.94 kpc. The likelihood of finding an unrelated ULX within this region depends critically on the host galaxy properties and the surface density of ULXs. For the general galaxy population, the average number of ULXs ($L_X > 10^{39}$ $\rm erg\,s^{-1}$) is reported to be $0.88 \pm 0.05$ \citep{Kovlakas2020}. For a typical galactic disk of radius 15 kpc, the expected number of ULXs in a 1.94 kpc radius circle is $\sim (1.94/15)^2 \sim 0.016$, giving a 1.6\% chance probability using a Poisson statistics. However, this probability is significantly enhanced in star-forming regions or star-forming galaxies, where ULXs are substantially more prevalent \citep{Swartz2011ApJ, Feng2011}. In such active environments, the density of ULXs could be higher by a factor of 5 to 10, corresponding to a chance probability of approximately $8$--$15$\% to detect an unrelated ULX-like source within 10 arcseconds in a star-forming galaxy.

\RE{In addition, for even nearby FRBs, low-mass X-ray binaries (LMXBs) and high-mass X-ray binaries (HMXBs) may contribute to contamination. Using a similar method as above, and including the luminosity functions of LMXBs, HMXBs, and ULXs \citep{Wangs2016, Gilfanov2022}, we estimate the average number of unrelated LMXBs, HMXBs, and ULXs expected within the angular resolution of EP/FXT (10$^{\prime\prime}$) and Chandra (1$^{\prime\prime}$) for different luminosity distances $D_\mathrm{L}$, as shown in Figure \ref{fig:Nave}.
It shows that the chance of contamination from LMXBs, HMXBs, and ULXs is about 1\% for EP/FXT and 0.01\% for Chandra.}
Furthermore, given that around 20\% FRBs reside in host galaxies with AGN signals \citep{Eftekhari2023, Sharma2024}, the contamination from AGNs or intermediate-mass black holes (IMBHs) is then non-negligible, especially for higher redshifts. These considerations highlight the importance of arcsecond-level localization for both FRBs and their potential X-ray counterparts, although EP/FXT candidates remain useful in most cases.

\subsection{One-off or Repeating}

\begin{figure*}
\centering
\begin{tabular}{c}
\includegraphics[width=0.45\textwidth]{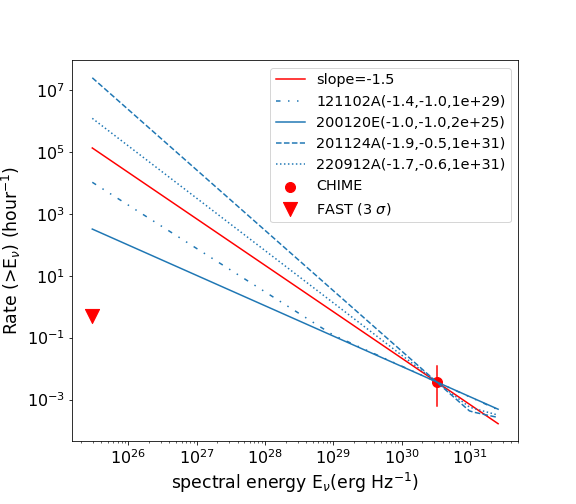}
\includegraphics[width=0.45\textwidth]{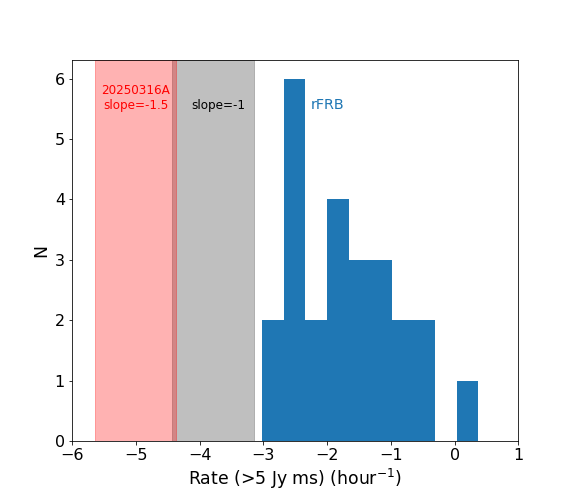}
\end{tabular}
\caption{Left: CHIME FRB rate and the FAST 3$\sigma$ upper limit compared with the spectral energy functions of repeating FRBs. A typical power law with a slope of $-1.5$ is plotted as a red solid line. The blue lines present the spectral energy functions of repeating FRBs, scaled to match the rate of FRB~20250316A at its spectral energy. Right: The FRB rate ($>$ 5 Jy ms) of FRB~20250316A estimated with multiple telescopes, assuming energy function slopes of $-1.5$ (red region) and $-1.0$ (gray region), compared with the CHIME-detected FRB rate (blue histogram). }
\label{fig:frbrate}
\end{figure*}

One of the most critical questions in FRB research is whether all FRBs repeat. As a nearby FRB, FRB~20250316A provides a valuable opportunity to examine this issue. Observationally, it was detected by CHIME after 269.77 hours of exposures over six years \citep{FRB20250316Achime}, corresponding to an FRB \RE{Poisson} rate of 0.004 per hour with a 1$\sigma$ range of $0.0006-0.012$ per hour (\RE{$> 1.7 \times 10^3$ Jy ms}, corresponding to $> 3.2 \times 10^{30}$ erg Hz$^{-1}$). FAST conducted 13.08 hours of follow-up observations over approximately one week (Section \ref{sec:res_fast}). Assuming a Poisson distribution, this indicates a 3$\sigma$ upper limit of \RE{$<0.5~\mathrm{hour^{-1}}$} (\RE{$> 15.5$ mJy ms}, corresponding to $> 3.0 \times 10^{25}$ erg Hz$^{-1}$). The left panel of Figure \ref{fig:frbrate} presents FRB rates estimated with CHIME (red dot) and the upper limit from FAST (red triangle). For comparison, we overplot the energy functions of repeating FRBs. The energy functions of repeating FRBs are empirically fitted as a power law, with a slope of $\sim -1.5$, spanning a range from $-0.7$ to $-2.5$ \citep{Law2017, Gourdji2019}. A flatter high-energy component has sometimes been proposed for repeating FRBs, such as FRB~20121102A \citep{LiD2021, Hewitt2022}, FRB~20200120E \citep{Zhangsb2024}, FRB~20201124A \citep{Zhangyk2022RAA, Kirsten2024}, and FRB~20220912A \citep{OuldBoukattine2024}. \RE{Their energy functions are shown as blue lines, with the lower index, higher index, and the break spectral energy labeled.} The flatter component typically extends over $1-2$ orders of magnitude in spectral energy. A comparison between FRB~20250316A and repeating FRBs reveals two possible scenarios.  
First, the typical power law with a slope of $-1.5$ (red solid line) predicts an FRB rate \RE{of $2.7\times10^{5}\,\mathrm{hr^{-1}}$ above the FAST limiting fluence of $15.5\,\mathrm{mJy\,ms}$, more than} five orders of magnitude larger than the FAST 3$\sigma$ upper limit.
Second, scaling the energy functions of repeating FRBs to the rate of FRB~20250316A at its spectral energy (blue lines) shows that repeating FRBs predict rates 3–7 orders of magnitude higher than the FAST upper limit.

\RE{We note that repeating FRBs are usually clustered, with waiting times often described by Weibull distributions \citep{Oppermann2018, Oostrum2020, chime2020periodic, Rajwade2020, Kirsten2024}. The Weibull shape parameter $k$ characterizes clustering: bursts are clustered if $k < 1$, with typical values $k = 0.3$–$0.9$ \citep{Oppermann2018, Oostrum2020, Luo2020}. Using the simulation method of \citealt{Luo2020}, we obtained $3\sigma$ FRB rate upper limits of 2.0, 0.7, and 0.5 $\mathrm{hr^{-1}}$ (at FAST limiting fluence) for $k = 0.3$, 0.5, and 0.9, respectively—still 2–7 orders of magnitude below predictions from repeating FRB energy functions.}

On the other hand, some repeating FRBs have $\leq$ 10 bursts, with FRB rates of $10^{-3} - 2$ bursts per hour at a completeness of $0.5-10$ Jy ms \citep{chimerepeater2023}. Their energy functions cannot be examined in detail. In order to compare with them, we estimated the FRB rate of FRB~20250316A by combining observations from multiple telescopes. A cumulative FRB rate with multiple telescopes can be estimated as 
$$
R(>E_{\nu})=\frac{N}{\sum\limits_i t_\mathrm{obs,i}f(E_{\nu,i})/f(E_{\nu,0})},
$$
where $f(E_{\nu})$ is the energy function of FRBs, $i$ iterates over different telescopes, and $E_{\nu}$ is the threshold of each telescope. To compare with the CHIME repeating FRBs, we adopt $f(E_{\nu,i}) \propto E_{\nu}^{\alpha}$ following \cite{chimerepeater2023}. \RE{Taking into account the observations of FAST, CHIME, and HyperFlash---a 25–32 m European radio telescope network that observed FRB~20250316A for 120 hours at L-band with a limiting fluence of 15 Jy ms \citep{Ould-Boukattine2025, FRB20250316Achime}---}and assuming a slope of $-1.5$, we obtain a combined FRB~20250316A rate of $(1.3^{+3.0}_{-1.0}) \times 10^{-5}$ hour$^{-1}$ (red region in the right panel of Figure \ref{fig:frbrate}). This region lies below all CHIME repeating FRBs (blue histogram). Using the flattest slope of well-studied repeating FRBs, $\sim -1.0$ (FRB~20200120E, \RE{\citealt{Zhangsb2024}}), the rate is estimated to be $(2.2^{+5.0}_{-1.0}) \times 10^{-4}$ hour$^{-1}$ (gray region), still below all CHIME repeating FRBs. Thus, FRB~20250316A is different from the currently known repeating FRB sample.

\RE{\citealt{FRB20250316Achime} explored the repeatility of this burst by a simulation, and concludes that the probability of detecting only one burst like FRB~20250316A in repeating FRBs is more significant than 5$\sigma$. With different methods, we have the same conclusion that}
FRB~20250316A is highly likely to be a one-off event. 
We caution, however, that this process assumes the energy function of repeating FRBs is stable, whereas temporal variations are possible \RE{\citep{LiD2021, Zhangsb2024}}. In addition, we use only the ``gold'' sample from \cite{chimerepeater2023}. The ``silver'' sample, which may be contaminated by chance coincidence, may yield CHIME repeating FRBs with a lower rate. \RE{As a result, it is hardly to totally exclude the possibility that FRB~20250316A would be detected in some distant future.}

\subsection{Constraints on multiwavelength counterparts}

Different kinds of X-ray counterparts are expected for FRBs. Prompt X-ray pulses with durations ranging from milliseconds to seconds are anticipated in the magnetar framework. Such counterparts were detected in association with bright radio pulses from the Galactic magnetar SGR 1935+2154 \citep{Mereghetti2020ApJ, Li2021NatAs, sgr1935atel2022}. Our observations do not cover the time of FRB~20250316A itself, so we do not focus on this case in detail. 

Persistent X-ray emission is also expected in various models.
In the magnetar scenario, outbursts of persistent X-ray emission are usually observed during periods of enhanced bursting and flaring activity \citep{Kaspi2017ARA&A, CotiZelati2018, Rea2025}, as in the case of the X-ray burst from the Galactic magnetar SGR~1935+2154 associated with FRB~20200428D \citep{STARE2, CHIME2020Natur}. Typical long-term X-ray outbursts of magnetars might last several months to years, reaching a maximum luminosity of $\sim10^{36}$\,erg s$^{-1}$ \RE{\citep{Kaspi2017ARA&A}, where most of the luminosity is expected to be released in neutrinos \citep{PonsRea2012}}. Given our upper limit of $\sim 10^{39}$ erg s$^{-1}$, we cannot place meaningful constraints on such persistent magnetar outbursts. However, if FRBs are produced by a very young magnetar with fast rotation, one might expect a large amount of rotational power ($E_{\rm rot}$) that could support an X-ray counterpart in the form of pulsed synchrotron emission or a wind nebula. For example, by assuming: 1) a magnetic field of at least $\sim10^{14}$\,G (required for the magnetar to show large X-bursts related to the FRB radio counterpart\RE{, \citealt{Turolla2015}}), and 2) an X-ray efficiency of $\epsilon_X \sim10^{-3}$ (in line with rotationally powered young pulsars\RE{, \citealt{Possenti2002}}), our deep upper limit implies a maximum rotational power for the young magnetar of $E_{\rm rot} < 10^{42}$\,erg s$^{-1}$ (assuming a NS mass of 1.4 $M_{\odot}$ and radius of 12\,km). Under these assumptions, the putative young magnetar responsible for FRB~20250316A would have a spin period larger than P $>50\,\mathrm{ms}$ and a characteristic age longer than 1~yr,  possibly in line with the magnetar population of this star-forming galaxy.

Recent FRB environment studies and possible periodicities imply a possible link to binary systems \citep{Ioka2020, Wangfy2022,Anna-Thomas2023,Liy2025, zhangb2025}. Thus, neutron star binary systems, especially LMXB, HMXBs and ULXs, might also be related to FRBs.

\begin{figure*}
\centering
\begin{tabular}{c}
\includegraphics[width=0.41\textwidth]{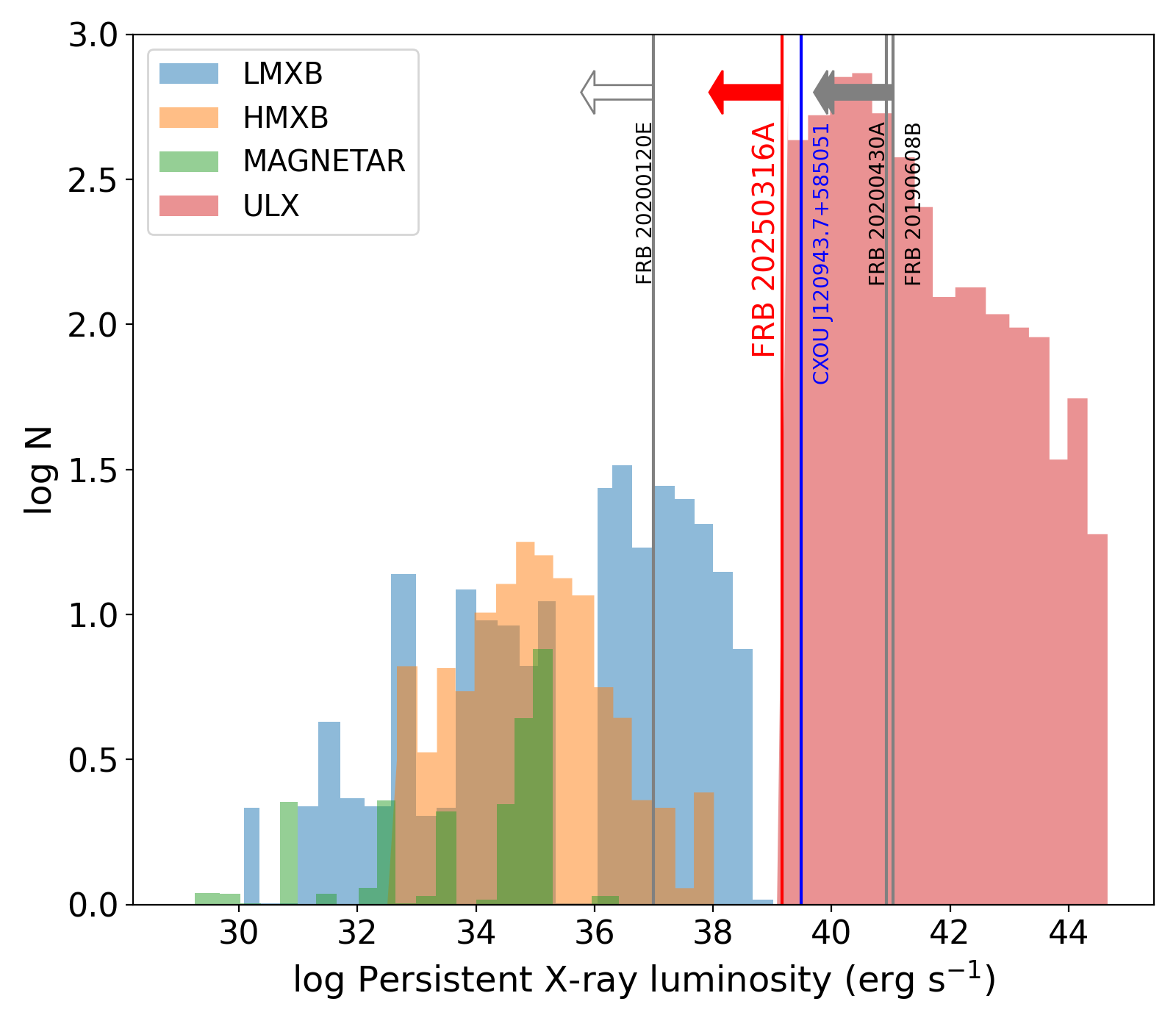}
\includegraphics[width=0.43\textwidth]{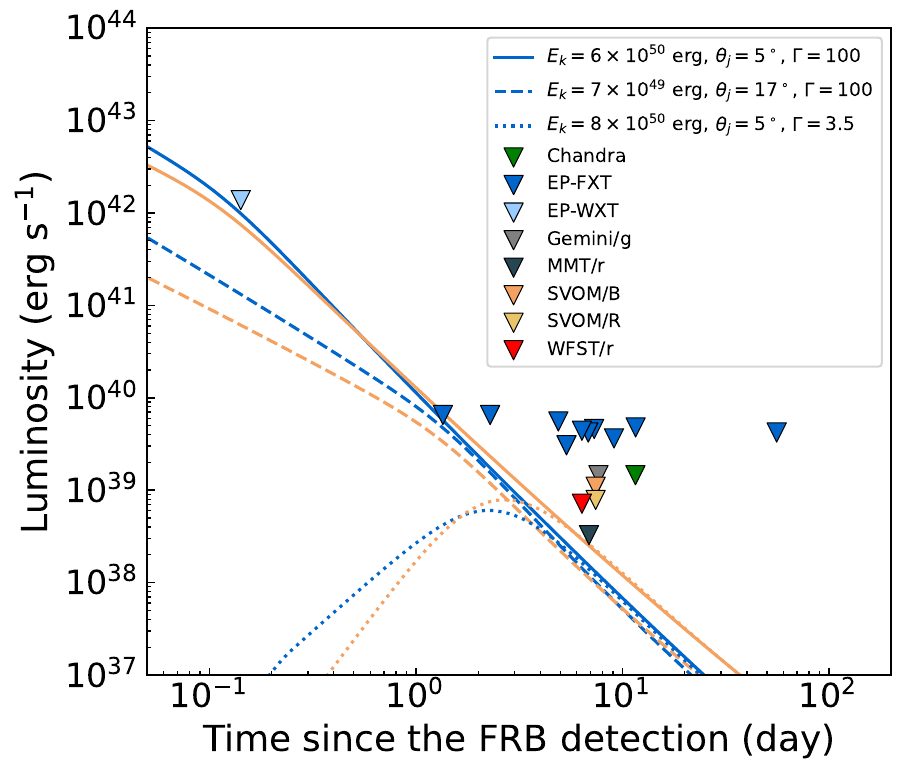}
\end{tabular}
\caption{Left: Upper limits on the persistent X-ray luminosity from the Chandra observation at the position of FRB~20250316A (red arrow). For comparison, the observed luminosities of LMXBs, HMXBs, magnetars and ULXs are overplotted. Upper limits on the persistent X-ray luminosity from other one-off FRBs (filled arrows) and repeating FRB (open arrow) are also shown. 
Right panel: Multiwavelength constraints of FRB~20250316A. The predictions of afterglow models \citep{vegas2025} with different parameters are overplotted. Blue and orange lines are for X-ray and optical, respectively.}
\label{fig:constrain}
\end{figure*}

In Figure \ref{fig:constrain}, we compare the persistent X-ray luminosity upper limit of FRB~20250316A with those expected from different neutron star classes, both in isolation and in binary systems. \RE{We also overplot upper limits from other deep X-ray observations, including the repeating FRB~20200120E (gray open arrow; \citealt{Pearlman2025}), as well as the non-repeating FRBs~20200430A and 20190608B (gray filled arrows; \citealt{Eftekhari2023}).} The Chandra X-ray upper limit of FRB~20250316A is fainter than nearly all ULXs, while still compatible with an X-ray binary or a magnetar. \RE{It is deeper than previously reported limits for other non-repeating FRBs, but less constraining than the limit for the repeating FRB~20200120E.} Although the strong variability observed in some ULXs prevents us from completely ruling out these systems as possible progenitors of FRB~20250316A, this result disfavors a direct connection between non-repeating FRBs and ULXs, \RE{in agreement with previous findings for the repeating FRB~20200120E}.

On the other hand, if FRBs are accompanied by outflows, multiwavelength afterglows are expected, typically lasting for days \citep{Yisx2014}.
In the right panel of Figure~\ref{fig:constrain}, we present the X-ray upper-limit constraints obtained from EP-WXT, EP-FXT, and \textit{Chandra}, together with the optical upper-limit constraints from SVOM and WFST for FRB~20250316A. Here, the 3$\sigma$ values of EP~J120944.2+585060 are adopted as upper limits of the FRB. Based on the X-ray upper-limit constraints, and assuming the source is viewed on-axis (which yields the strongest constraint on the kinetic energy), three possible scenarios can be considered.
\begin{itemize}
\item {\bf Ultra-relativistic outflow}:
If the fireball decelerates before the first WXT observation (i.e., for $\Gamma \gtrsim 100$), the WXT point at $\sim 0.1$ day provides the most stringent kinetic energy constrain, $E_k \lesssim 6 \times 10^{50}$ erg. The first FXT point at $\sim 1$ day then limits the jet opening angle to $\sim 5^\circ$, requiring a jet break before $\sim 0.5$ day to keep the X-ray flux below the FXT measurements (see blue solid line). The corresponding optical emission (orange solid line) is also consistent with the upper limits from SVOM and WFST, as well as the absence of radio counterparts \citep{Ant2025}.

\item {\bf Moderately wide jet}:  
For a larger jet opening angle of $\theta_j \sim 17^\circ$, the jet break occurs near the first FXT observation. In this case, the WXT data no longer provide the strongest constraint. Instead, the first FXT point sets the dominant limit, yielding $E_k \lesssim 7 \times 10^{49}$ erg. The corresponding X-ray and optical lightcurves are shown by the blue and orange dashed lines.  

\item {\bf Low Lorentz factor outflow}:  
If the fireball decelerates around $\sim 1$ day (near the FXT points), the inferred Lorentz factor is relatively low, $\Gamma \sim 5$. In this scenario, the FXT upper limits do not constrain the jet kinetic energy. Instead, the optical upper limits from SVOM and WFST
dominate, requiring $E_k \lesssim 1.2 \times 10^{51}$ erg. The corresponding X-ray lightcurve is shown as the blue dotted line.
\RE{If we take the MMT upper limit ($r>25~\mathrm{mag}$) and and Gemini ($g > 23.7~\mathrm{mag}$ on 2025 March 24) into account, it requires $E_k \lesssim 8 \times 10^{50}$. }
\end{itemize}

These kinetic energy constraints are far larger than the typical FRB energy budget. Therefore, if a hypothetical FRB afterglow exists, the required radio efficiency, $\sim E_{\rm FRB}/E_k$, would have to be $\gtrsim 10^{-12}$. It is obvious that an early deep X-ray observation is important to further constrain the X-ray counterpart predicted by the afterglow model. 

\section{Summary}

In this paper, we present multiwavelength follow-up observations of the nearby bright apparent one-off FRB~20250316A, including radio observations with FAST, X-ray observations with EP and \textit{Chandra}, and optical observations with WFST and SVOM/VT. The non-detection of pulses with the FAST radio telescope indicates that \RE{the predicted rate of well-studied repeating FRBs at the FAST limiting fluence is $2$–$8$ orders of magnitude higher than FAST’s 3$\sigma$ upper limit, and that the effective FRB rate at the CHIME limiting fluence is lower than the rates of all known repeating FRBs.} These results suggest that this burst is distinct from the current sample of repeating FRBs and is likely a one-off event, \RE{supporting the results of \citealt{FRB20250316Achime}.} The EP-WXT obtained the first upper limit of  $7.2\times 10^{-12}\,\rm\,erg\,cm^{-2}\,s^{-1}$ at just 3.5 hours post-burst, followed by the EP-FXT achieving a much deeper limit of $ 3.4\times 10^{-14}$ $\rm erg\,cm^{-2}\,s^{-1}$ at 1.4 days after the FRB. \RE{The subsequent \textit{Chandra} observations established the most constraining 0.5--10 keV flux upper limit of $7.6\times 10^{-15}$ $\rm erg\,cm^{-2}\,s^{-1}$, corresponding to an X-ray luminosity upper limit of $\sim 10^{39}$ $\rm erg\,s^{-1}$, which is lower than those of most ULXs.} These results place one of the most stringent limits on persistent X-ray emission from a non-repeating FRB, disfavoring the association of FRBs with ULXs and providing critical constraints for theoretical models of FRB progenitors and emission processes. In addition, the combination of the optical and X-ray upper limits constrain the kinetic energy of afterglow models to $\rm \lesssim 10^{51}~erg$. 

Based on the chance coincidence probability estimation, we should be cautious about future FRB multiwavelength counterpart searches: arcsecond-level localization of both FRBs and their potential X-ray counterparts is essential for a convincing counterpart search. 

\begin{acknowledgments} 
This work is based on data from Einstein Probe, a space mission supported by the Strategic Priority Program on Space Science of the Chinese Academy of Sciences, in collaboration with the European Space Agency (ESA), the Max-Planck-Institute for Extraterrestrial Physics, and the Centre National d'Études Spatiales (CNES) (Grant No. XDA15310000).
\RE{This research uses Chandra datasets from the Chandra X-ray Observatory~\dataset[DOI: 10.25574/30871]{https://doi.org/10.25574/30871}.}
FAST, a Chinese national mega-science facility, is built and operated by the National Astronomical Observatories, CAS.
WFST is jointly operated by the University of Science and Technology of China and Purple Mountain Observatory.
The Space-based multi-band Variable Objects Monitor (SVOM) is a joint Chinese-French mission led by the Chinese National Space Administration (CNSA), CNES, and CAS.

This work is supported by the Strategic Priority Research Program of the Chinese Academy of Sciences (Grant No. XDB0550200), the National Natural Science Foundation of China (Grant Nos. 12041306, 12321003, 12103089, 12393813, 12025303, 12041303, 12421003), the National Key Research and Development Program of China (2022SKA0130100,2020SKA0120200), International Partnership Program of Chinese Academy of Sciences for Grand Challenges (114332KYSB20210018), the CAS Project for Young Scientists in Basic Research (Grant No. YSBR-063), the CAS Organizational Scientific Research Platform for National Major Scientific and Technological Infrastructure: Cosmic Transients with FAST, the Young Elite Scientists Sponsorship Program by China Association for Science and Technology (Grant No. YESS20240218), and Beijing Nova Program (No. 20250484786).

\end{acknowledgments}

\appendix
\renewcommand{\thetable}{A\arabic{table}} 
\setcounter{table}{0} 
\begin{threeparttable}
\begin{center}
\scriptsize
\setlength{\tabcolsep}{2 pt}
\begin{longtable}{cccccc}
\caption{\RE{Example of Optical Observations and Upper Limits}}
\label{tb:opt1img}\\
\hline
No & telescope & start time & MJD & filter & mag$_\mathrm{lim}$ \\
\hline
\endfirsthead
\multicolumn{6}{c}{{\tablename\ \thetable{} -- Continued}} \\
\hline
No & telescope & start time & MJD & filter & mag$_\mathrm{lim}$ \\
\hline
\endhead
\hline
\multicolumn{6}{r}{{Continued on next page}} \\
\endfoot
\hline
\endlastfoot
1 & WFST & 2025-03-22 17:04:19 & 60756.71133 & $r$ & 23.12 \\
\hline
\hline
\end{longtable}
Only the first few rows are shown here; the full table is available in the online version as a supplementary file.
\end{center}
\end{threeparttable}


\bibliographystyle{aasjournalv7}




\end{document}